\documentclass[amsmath,amssymb,floatfix,prb,epsf,showpacs,twocolumn,graphicx]{revtex4}
\usepackage{amsmath,amssymb,natbib,bm,graphicx,url,epsfig}
\usepackage[ansinew]{inputenc}
\usepackage{bm}

\usepackage{graphicx}
\usepackage{epstopdf}

\newcommand{\be}{\begin{equation}}
\newcommand{\ee}{\end{equation}}
\newcommand{\bea}{\begin{eqnarray}}
\newcommand{\eea}{\end{eqnarray}}
\newcommand{\bes}{\begin{equation}\begin{split}}
\newcommand{\ees}{\end{split}\end{equation}}

\begin{document}
\title{Quantum-critical pairing with varying exponents}

\author{Eun-Gook Moon}
\affiliation{Department of Physics, Harvard University, Cambridge MA
02138}

\author{Andrey Chubukov}
\affiliation{
Department of Physics, University of Wisconsin, Madison, Wisconsin 53706}

\date{\today}

\begin{abstract}
We analysed the onset temperature $T_p$ 
for the pairing in cuprate superconductors  at small doping, when tendency
 towards antiferromagnetism is strong.  We considered the model of
 Moon and Sachdev (MS), which assumes that electron and hole pockets
survive in a paramagnetic phase. Within this model, 
the pairing between fermions is mediated by a gauge boson, whose propagator remains massless in a paramagnet.  We related the MS model to a generic $\gamma-$model 
of quantum-critical pairing with  the pairing kernel $\lambda (\Omega_n) \propto 1/\Omega^{\gamma}_n$. We showed that, over some range of parameters,
 the MS model is equivalent to $\gamma =1/3$-model ($\lambda (\Omega) \propto \Omega^{-1/3})$. We found, however, 
 that the parameter range where this analogy works is bounded on both ends. 
At larger deviations from a magnetic phase, the MS model becomes equivalent to $\gamma$ model with varying $\gamma >1/3$, whose value depends on the distance to a magnetic transition and approaches $\gamma =1$ deep in a paramagnetic phase. 
Very near the transition, the MS model  becomes equivalent to $\gamma$ model with varying $\gamma <1/3$. Right at the magnetic QCP, the MS model is equivalent to 
 the model with $\lambda (\Omega_n) \propto \log \Omega_n$, which is the model for color superconductivity. Using this analogy, we verified the formula 
for $T_c$ derived for color superconductivity.
\end{abstract}

\maketitle

\section{Introduction}
\label{sec:intro}

A quest for understanding of the  phase diagram of cuprates and other strongly correlated electron systems generated strong interest to 
  the pairing problem near a quantum critical point (QCP). 
In distinction to a conventional 
BCS/Eliashberg theory, in which pairing is mediated by an interaction which can be approximated by a constant at small frequencies,
 quantum-critical pairing is mediated by gapless, dynamical collective modes~\cite{acs}. 
In most cases, only a particular charge or spin mode becomes gapless at criticality, and the same interaction that gives rise to pairing
 also accounts for the fermionic self-energy. In systems with $d \leq 3$,  $\partial \Sigma(k, \omega)/\partial \omega$ diverges at $\omega \to 0$, i.e., interaction with a gapless collective modes destroys fermionic coherence  either at particular hot spots~\cite{acs}, if the energing order parameter has a finite momentum $q$, or along the whole Fermi surface (FS) if the ordering is with $q=0$ (Ref.\cite{q=0}).  Then, in another distinction to a conventional BCS/Eliashberg theory, the pairing involves incoherent fermions.

 The self-energy at a QCP generally behaves as $\omega_0^\gamma \omega^{1-\gamma}$, where $\gamma >0$, while the effective dynamic pairing interaction (the analog of phonon 
$\alpha^2 (\Omega) F(\Omega)$) scales $\lambda (\Omega) \propto (\Omega_0/\Omega)^{\gamma}$. 
The kernel $K(\omega,\omega')$ of the equation for the pairing vertex  
$\Phi (\omega) = \pi T \sum_{\omega'} K(\omega, \omega') \Phi (\omega')$  then scales as
\be
K (\omega, \omega') = \frac{\lambda (\omega-\omega')}{|\Sigma (\omega')|}
 \propto \frac{1}{|\omega -\omega'|^{\gamma} |\omega'|^{1-\gamma}}.
\label{1a}
\ee
It has the same scaling dimension $-1$ as in BCS theory, which
 corresponds to $\gamma =0$.  However, in distinction to BCS theory, where the kernel is  $1/|\omega'|$, the kernel  $K(\omega - \omega')$ for $\gamma >0$ 
 depends on both external and internal energy, and the onset temperature for the pairing $T_p$ has to be obtained by solving integral equation for the frequency-dependent pairing vertex $\Phi (\omega)$ (we label this temperature as $T_p$ to distinguish it from the actual superconducting $T_c$ which is generally lower because of phase and amplitude fluctuations of  $\Phi (\omega)$). 
The frequency dependence  of the vertex by itself is not unique to a
 QCP and is present already in the Eliashberg theory of conventional superconductors~\cite{phon_rev}.
 However, as we said, in conventional cases, the frequency dependence is relevant at high frequencies, while the low-frequency sector can still be treated 
 within BCS theory. The new element of quantum-critical pairing is that non-trivial frequency dependence of the pairing vertex 
  extends down to $\omega =0$ leaving no space for  the BCS regime. 

To shorten the notations, below we label the model
 of QC pairing mediated by $\lambda(\Omega) \propto 1/|\Omega|^\gamma$ as
 the $\gamma$-model. Examples of $\gamma-$models, include 
 2D pairing by ferromagnetic or antiferromagnetic spin fluctuations 
($\gamma \approx 1/3$ and $\gamma \approx 1/2$, respectively~\cite{acs,acf,q=0,max}, and 3D pairing by gapless charge or spin fluctuations ($\gamma = 0+$, 
implying that $\lambda (\Omega) \sim \log \Omega$  and $\Sigma (\omega) \sim \omega \log \omega$, Ref.~\cite{bedell,ch_sch}). 
Another example of $\gamma = 0+$ behavior is color superconductivity of quarks mediated by gluon exchange~\cite{son}.  The $\gamma =1/3$ model describes the pairing of composite fermions at $\nu =1/2$ Landau level.~\cite{nick} 
A similar problem with momentum integrals instead of frequency integrals have been considered in Ref. \onlinecite{khvesh}.  

The solutions of the pairing problem in 
 these systems differ in details but have one key feature in common --
 $T_p$ monotonically increases upon approaching a QCP from a disordered side.
At a QCP, $T_p = \Omega_0 ~f(\gamma)$, where $f(\gamma)$ is a smooth function,
 which decreases when $\gamma$ increases.  

Recently, Moon and Sachdev (MS) considered~\cite{moon} another example of 
 QC pairing --  the pairing of fermions in the presence of  spin-density-wave (SDW) 
 background, which can be  either in the form of  long-range SDW order, 
or SDW precursors.
MS argued that the phase with SDW precursors is an algebraic charge-liquid
 in which fermions still posssess the same pocket FS as in the SDW phase~\cite{subir}.  
They found that the model remains critical away from
critical point because of the presence of gapless gauge fluctuations, and 
$T_p$ actually increases as the system moves away from a QCP
 into a charge-liquid phase. This trend is opposite compared to a 
 magnetically-mediated pairing without SDW precursors. 
This result is quite 
 important for the understanding of the pairing in the cuprates as it shows that  the onset temperature of the pairing instability  does form a dome centered around a doping  at which the system develops SDW precursors.~\cite{subir,sachdev,sedr}  In the region where SDW precursors are not yet formed, there is a large
 Fermi surface, and $T_p$ increases as doping decreases and the system comes closer to a magnetic instability. Once SDW precursors are formed above $T_p$, the trend changes, and $T_p$ passes through a maximum and 
 begins decreasing with decreasing doping.  

The reason for the opposite behavior of $T_p$ in the presence of SDW precursors
 is  the suppression of the $d-$wave pairing vertex. 
 In the ordered SDW state, the  interaction between fermions mediated by Golstone bosons is dressed up by SDW coherence factors  and vanishes at the ordering momentum by Adler principle~\cite{SDW_ordered} 
 In the disordered phase, there are no such requirement, but the vertex is still suppressed as long as the system  has SDW precursors. As the system moves away  from a QCP, SDW precursors weaken, the pairing interaction gradually increases towards its value without precursors, and $T_p$ increases.

The goal of this communication is to relate the pairing problem in the presence of SDW precursors with earlier studies of the pairing within the $\gamma$ model. We argue that the MS model belongs to the class of $\gamma$-models,
  however the relation between the two is rather non-trivial.
 First, the pairing in the MS model remains quantum-critical even 
away from the QCP, when spin excitations acquire a finite mass $m$
 at $T=0$, i.e., the MS model away from the QCP is equivalent to a 
$\gamma$-model right at criticality.  Second, we show that the MS model at different deviations from the QCP is equivalent to  $\gamma-$models with different $\gamma$, 
  ranging from  $\gamma = 0+$ right at the QCP to $\gamma =1$ deep into the
magnetically-disordered phase (but still with SDW precursors). Third, the
 scale $\Omega_0$ in $\lambda(\Omega) = (\Omega_0/|\Omega|)^\gamma$ also  becomes $\gamma$-dependent and increases with increasing $\gamma$. This
 is the key reason why $T_p \propto \Omega_0$ increases into the disordered state. 
       
We also consider in more detail the pairing problem right at the QCP. We find that at this point,  the pairing interaction at small frequencies is $\lambda (\Omega) = g \log (\Omega_0/|\Omega|)$
 and the fermionic self-energy  $\Sigma (\omega) = g \omega \log (\Omega_0/|\omega|)$, are logarithmical and  
depend separately on the cutoff scale  $\Omega_0$ and on the dimensionless 
coupling constant $g$.  This separate dependence is unique to the logarithmical form of the pairing interaction, for any power-law form of $\lambda (\Omega) = (\Omega_0/\Omega)^{\gamma}$, the coupling $g$ is incorporated into $\omega_0$. 
 At frequencies $\Omega > \Omega_0$, $\lambda (\Omega)$ decays faster than the logarithm, eventially as $1/\Omega$. As we said, the same logarithmical form of the pairing vertex appears in the problem of color superconductivity. For 
 color superconductivity, Son obtained at weak coupling $T_p \propto \omega_0 e^{-\pi/(2\sqrt{g})}$ [Ref.\cite{son}]. The result is similar to the BCS formula, but the dependence on the coupling is $\sqrt{g}$ rather than $1/g$. Son's result was confirmed in 
subsequent analytical studies, but, to the best of our knowledge, it has not been verified by  actualy solving the linearized gap equation for the 
 case when the interaction is logarithmical at small frequencies and decays as a power of $1/\Omega$ at larger frequencies. We solved this equation for our $\lambda (\Omega)$, obtained $T_p$, and found excellent agreement with $e^{-\pi/(2\sqrt{g})}$ behavior.      

The paper is organized as follows. In Sec. \ref{sec:2} we briefly review 
 (i) Eliashberg-type theory for the pairig induced by frequency dependent pairing kernel $K(\Omega)$, (ii) the $\gamma$-model with no SDW precursors, and (iii) the MS model for the pairing in the presence of SDW precursors.
 In Sec. \ref{sec:3} we show that the MS model reduces to the set of
 $\gamma-$models with  varying $\gamma$, which gradually increases from $\gamma = 0+$ as one moves away from the QCP. We argue that $\gamma$ remains $1/3$ over some range of deviations from the QCP, but this range is quite narrow.
 In Sec. \ref{sec:4}
 we consider in more detail the MS model at the QCP and show that at this point it is equivalent to the $\gamma$ model at $\gamma = 0+$ which in turn is equivalent to the model for color superconductivity. We present the numerical solution for the pairig vertex, and show that $\log T_c/E_F$ scales as $1/\sqrt{g}$. 
 Sec. \ref{sec:5} prsents our conclusions. 

\section{Review of the models}
\label{sec:2}

\subsection{Eliashberg theory}

Throughout the paper we assume that the pairing problem can be treated within 
 Eliashberg-type theory, i.e., assume that
 the pairing kernel and the fermionic self-energy
 have singular frequency dependence but no substantial momentum dependence. 
The justifications for neglecting the momentum dependence are to some extent problem-specific, but generally are due to the fact that collective bosons are Landau overdamped what makes them  slow modes compared to electrons.

Within Eliashberg theory, the fermionic Green's function  in Nambu  spinor notations is given by
\begin{eqnarray}
{\hat \Sigma} (  \omega_{n}) & = & i \Sigma (\omega_{n}) \hat \tau_{0} + \Phi( \omega_{n}) \hat \tau_{1} \\
{\hat G}^{-1}(\epsilon_k,  \omega_{n})& = & i (\omega_{n} + \Sigma ( \omega_{n})) \hat \tau_{0} - \epsilon_k \hat \tau_{3} - \Phi( \omega_{n}) \hat \tau_{1} \nonumber
\label{1}
\end{eqnarray}
where ${\hat G}^{-1} = {\hat G}_0^{-1} + {\hat \Sigma}$, 
 $\hat{\tau}$ are the Pauli matrices in the particle-hole space, $\epsilon_k$ is the fermionic dispersion in the normal state, $ \Phi( \omega_{n})$ is a pairing vertex, and $\Sigma (\omega_n)$ is a conventional (non-Nambu) 
self-energy. 
The Nambu  self-energy $\hat \Sigma (  \omega_{n})$ 
in turn is expressed via the full Green's function as
\begin{eqnarray}
 \hat \Sigma ( i \omega_{n})  &=& - T \sum_{\omega_m} \lambda (\omega_{m}-\omega_{n}) \int d \epsilon_k \hat G(\epsilon_k,  \omega_{m}) \nonumber \\
 & = & i \Sigma (\omega_{n}) \hat \tau_{0} + \Phi( \omega_{n}) \hat \tau_{1} \nonumber
\label{2}
\end{eqnarray}

The full set of  Eliashberg equations is
 obtained by matching the coefficients of the Pauli matrices term by term:
\begin{eqnarray}
\Sigma (\omega_n) &=& \pi T \sum_{\omega_m } \lambda (\omega_m- \omega_n) 
\frac{\omega_{m} + \Sigma (\omega_n)}{\sqrt{(\omega_{m} + \Sigma(\omega_m))^2  + \Phi^2( \omega_m)}} \nonumber\\
\Phi (\omega_n) &=& \pi T \sum_{\omega_m } \lambda (\omega_m-\omega_n) \frac{  \Phi( \omega_{m})}{\sqrt{(\omega_{m} + \Sigma(\omega_m))^2  + \Phi^2( \omega_m)}} \nonumber \\
\label{Eliashberg}
\end{eqnarray}
This set of two non-linear self-consistent integral equations can be reduced
 to just one integral equation by introducing the inverse quasiparticle 
 renormalization factor $Z (\omega_n) = 1 + \Sigma (\omega_n)/\omega_n$ and the pairing gap $\Delta (\omega_n) = \Phi (\omega_n)/Z(\omega_n)$.  The equation for $\Delta (\omega_n)$ then decouples from the equation on $Z(\omega_n)$ 
 and takes the form
 \begin{equation}
\Delta (\omega_n) = \pi T \sum_{\omega_m } \frac{\lambda (\omega_m-\omega_n)}
{\sqrt{\omega^2_{m}  + \Delta^2( \omega_m)}} \left(\Delta( \omega_{m}) - \Delta (\omega_n) \frac{\omega_m}{\omega_n}\right)
\label{3}
\end{equation}
The inverse quasiparticle renormalization factor $Z(\omega_n)$ is then obtained by substituting the result for $\Delta (\omega_m)$ into 
 \begin{equation}
\omega_n Z (\omega_n) = \omega_n +  \pi T \sum_{\omega_m }
\lambda (\omega_m-\omega_n) \frac{\omega_{m}}{\sqrt{\omega^2_{m} +
 \Delta^2 (\omega_m)}}
\label{4}
\end{equation}     

To obtain $T_p$ we will need to set $\Phi, \Delta \to 0$ and solve the linearized equation either 
for the pairing vertex  or the pairing gap. Both are non-self-consistent equations in this limit
 because at $\Phi$ can be safely dropped from the expression for 
 $\Sigma (\omega_n)$ which becomes:
\begin{equation}
\Sigma (\omega_n) = \pi T \sum_{\omega_m } 
\lambda (\omega_m- \omega_n) {\text sign} (\omega_m), ~~Z(\omega_n) = 1 + \frac{\Sigma (\omega_n)}{\omega_n}
\label{5}  
\end{equation}
It is more straighforward to analyze the linearized 
equation for the pairing vertex $\Phi (\omega_n)$:
 \begin{equation}
\Phi (\omega_n) = \pi T \sum_{\omega_m } K(\omega_n, \omega_m) \Phi (\omega_m),
\label{6}  
\end{equation}  
where the kernel is
\begin{equation}
K (\omega_n, \omega_m) = \frac{\lambda (\omega_m- \omega_n)}{|\omega_m| Z(\omega_m)} 
\label{7}
\end{equation}

\subsection{the $\gamma$-model}

We use the term $\gamma$-model as abbreviation for a critical model
 in which the pairing is mediated by a gapless boson, 
 and  $\lambda (\omega_n-\omega_m)$ has a power-law form 
\begin{equation}
\lambda (\Omega) = \left(\frac{\Omega_0}{|\Omega|}\right)^\gamma
\label{8}
\end{equation}
We assume in this work that $\gamma <1$. The analysis of the $\gamma-$models  with $\gamma \geq 1$ is more involved (and more non-trivial), but we will only need $\gamma < 1$ for comparisons with the MS theory.

To put this into perspective, we remind that in  BCS theory $\lambda (\Omega_n)$  is a constant up to a cutoff scale above which it is set to zero. In Eliashberg theory for non-critical superconductors 
 $\lambda (\Omega_n)$ has some frequency dependence but still reduces to a constant at small frequencies. 
In this situation, the pairing problem differs from BCS problem quantitatively but not qualitatively. In Eq. (\ref{8}), however, the pairing interaction 
 preserve sigular frequency dependence down to $\Omega =0$.
This makes the pairing problem in the $\gamma-$model qualitatively 
different from BCS problem.  

The divergence of $\lambda (\Omega_m)$  at zero bosonic Matsubara frequency $\Omega =0$ is by itself not relevant for the 
 pairing as can be straighforwardly seen from Eq. (\ref{3}) 
for $\Delta (\omega_m)$:  the term with $m=n$ in the r.h.s. of this equation vanishes. This vanishing is 
 essentially the  manifestation of the Anderson theorem for a dirty $s-$wave superconductors because the scattering with zero frequency transfer is formally 
analogous to impurity scattering~\cite{acf,msv,acn}. The recipe then is to
 eliminate the term with zero bosonic Matsubara frequency from the frequency sums, what we will do.

\begin{figure}
\includegraphics[width=3.0 in]{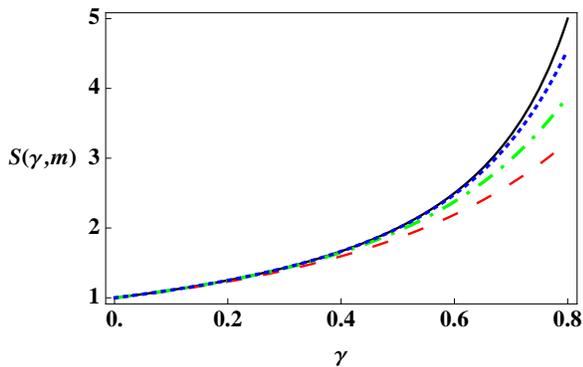}
\caption{ 
 Generalized Riemann-zeta function, $S(\gamma,m)$. The dashed (red), dot-dashed (green) and dotted (blue) lines correspond to $m= 100, 1000, 10000$. The black thick line is  the asymptotic function  $1/(1-\gamma)$ valid at large $m$ (see the text). }
\label{zeta}
\end{figure}

The fermionic self-energy for such $\lambda (\Omega)$ has the form
\begin{equation}
\Sigma (\omega_n) = \omega_n
 \left(\frac{\Omega_0}{|\omega_n|}\right)^\gamma  S(\gamma,n)
\label{9}
\end{equation}
where 
\be
S(\gamma,m) = \frac{1}{(m+1/2)^{1-\gamma}} \left[\zeta(\gamma)-\zeta(\gamma,m+1)\right] 
\ee
where $\zeta (\gamma)$ and $\zeta (\gamma, m+1)$ are Rieman zeta function and generalized Rieman zeta function, respectively. 
 We plot $S(\gamma,m)$ in Fig. \ref{zeta}.  At large $m$, $S(\gamma, m)$  approaches $1/(1-\gamma)$, 
 and the self-energy becomes
 $\Sigma (\omega_n) = \omega^{1-\gamma}_n (\Omega^{\gamma}_0/(1-\gamma)$  
(this limiting form is reproduced if the summation over $\omega_m$ is
 replaced by the integration).

 The rate with which $S(\gamma, m)$ approaches $1/(1-\gamma)$ by itself depends on $\gamma$ and slows down when $\gamma$ approaches one. For $1-\gamma <<1$, $S(\gamma,m)$ at large $m$ behaves as 
$S(\gamma,m) \approx (1- m^{-(1-\gamma)})/(1-\gamma)$ and becomes $\log m$ at $\gamma =1$.

Substituting the self-energy into the equation for $\Phi (\omega_m)$ and rescaling the temperature by $\Omega_0$ we find, instead of (\ref{6}),
 \begin{eqnarray}
&&\Phi (\omega_n) = \sum_{\omega_m } \frac{\Phi (\omega_m)}{|2m+1|} \times 
\nonumber \\
&& \frac{1}{|S(\gamma, m) \left(\frac{|m-n|}{|m+1/2|}\right)^\gamma + (2\pi |m-n|)^\gamma (\frac{T}{\Omega_0})^\gamma|}.
\label{10}  
\end{eqnarray}  
It is a'priori not guaranteed that this equation has a solution at some $T_p$,
 but, if it does, $T_p$ obviously scales as  
\be
T_p = \Omega_0 f(\gamma)
\label{11}
\ee
 Eq (\ref{10}) has been analyzed both analytically and numerically~\cite{acf,bedell,ch_sch,nick,khvesh,emil},
 and the result is that (i) the solution exists, and (ii) $f(\gamma)$ monotonically decreases with increasing $\gamma$, and remains finite at $\gamma =1$, where $f(1) = 0.254$ (Ref.\onlinecite{ch_sch}). 

\subsection{the MS model}

The MS model \cite{moon}
 describes a phase transition between an SDW state and a magnetically disordered, algebraic charge liquid which still preserves electron pockets near 
 $(0,\pi)$ and symmetry related points in the Brillouin zone. 
 The model is two-dimensional and involves non-relativistic fermions with small FS, relativistic bosons which 
describe antiferromagnetic magnons,      
 and a gauge field which couples bosonic and fermionic fields. 
There are two  non-relativistic fermions (two FSs), with the action
\begin{eqnarray}
\mathcal{L}_g &=& g_{\pm}^\dagger \left[ \partial_\tau   - \frac{{\bf \nabla} ^2}{2m^*}  - \mu  \right] g_{\pm}. \label{lg}
\end{eqnarray}
where $m^*$ is a fermionic mass, and the
 chemical potential $\mu$ is chosen to fix the Fermi momentum $k_F$.

The action of a relativistic boson 
 (an antiferromagnetic magnon) is
\begin{equation}
\mathcal{L}_z =   \sum_{\alpha=1}^{N} \biggl(|\partial_\tau  z_\alpha|^2 +
v^2 |{\bf \nabla}  z_\alpha|^2  + m^2 |z_\alpha|^2 \biggr)  \label{lz}
\end{equation}
 where $m$ is the mass of a magnon
\be
m=2 T \ln \left( {\frac{e^{\frac{2 \pi m^* v^2 \alpha_2}{T}}+ \sqrt{4+e^{\frac{4 \pi m^* v^2 \alpha_2}{T}}} }{2}} \right)
\label{18_a}
\ee
and the parameter $\alpha_2$ measures the distance to a QCP (we use the same notations as in~\cite{moon}). For
$\alpha_2 <0$ and $T=0$, the system is in the ordered SDW phase ($m=0$), for $\alpha_2 >2$ and $T=0$, the system is in the disordered phase ($m>0$). 
 At $T>0$, there is no long-range order, and $m>0$ for 
  all $\alpha_2$,  although for $\alpha_2 <0$ it is exponentially small.    

The gauge field couples bosonic and fermionic fields via minimal 
coupling$(\partial_{\mu} \rightarrow \partial_{\mu} \pm i  A_{\mu})$. 
The two fermions ($g_{\pm}$)  have opposite gauge charges, and gauge-field induced coupling between fermions and magnons is attactive and give rise to pairing. Both longitudinal and transverse parts of the interaction 
contribute  to the pairing, but non-trivial physics 
comes from the interaction mediated by a transverse boson. For simplicity, 
 we only consider transverse interaction.   

\begin{figure}
\includegraphics[width=2.5 in]{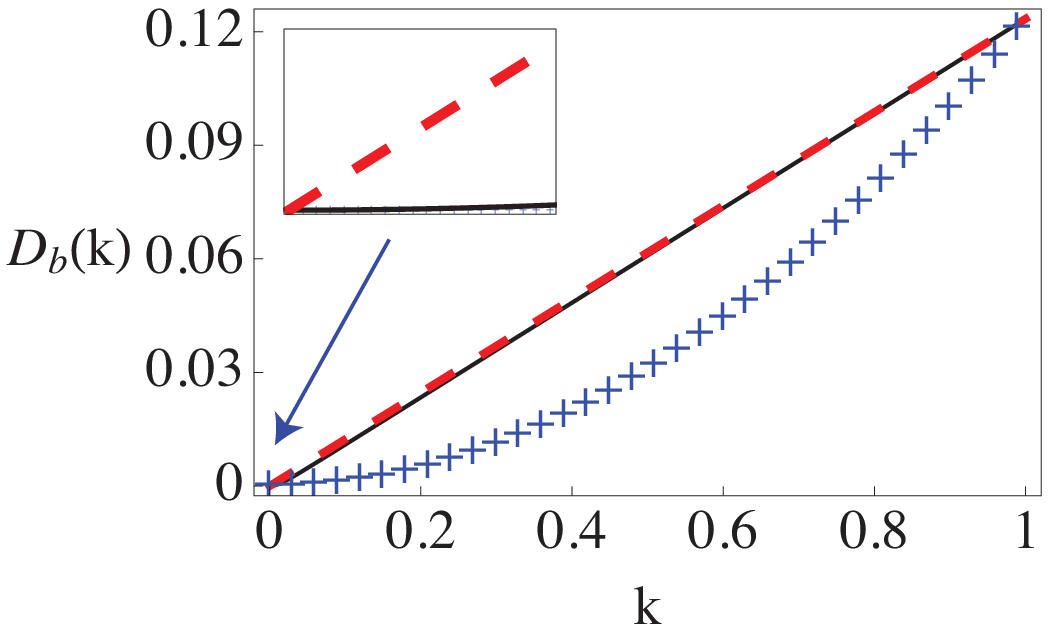}
\includegraphics[width=2.5 in]{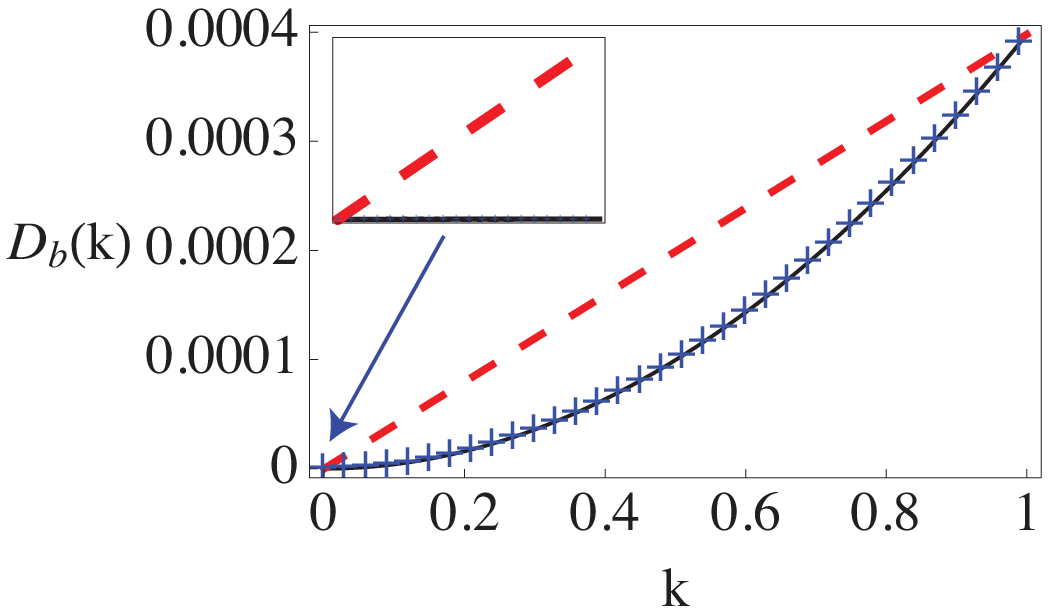}
\caption{
The momentum dependence of $D_b(k)$ at zero temperature (black line) 
 very near the QCP (top panel,  $m=0.01$) and far away from the QCP (lower panel,  $m=100$)
 $k$ is in units of $2k_F$
  and $D_b(k)$ is in units of $m^* v^2$ with $\alpha_1=0.5$. 
The thin-dashed line (Red) and the  thin-dotted line (Blue) 
show linear and quadratic behavior, respectively.  At the smallest $k$, 
$D_b (k)$ is quadratic for all $m >0$ (see inserts), but for small $m$ $D_b(k)$ raidly crosses over to the linear behavior. For large $m$, the quadratic dependence survives up to maximal $k=2k_F$.
 Note that the magnitude of $D_b (k)$ far away from the critical point is much smaller than near the QCP.}
\label{Db}
\end{figure}

The form of the transverse propagator of the gauge field 
 $\chi_{tr} (k, \Omega_n)$  is obtained  by combining the Landau damping term coming from fermions
 with the  static part coming from magnons. MS found, for the physical case of $N=2$ bosonic spinors, 
\begin{equation}
\chi_{tr} (k, \Omega_n) = \left( \delta_{ij} - \frac{k_i k_j}{k^2} \right) \frac{1}{2 D_{b} (k) + k_F |\Omega_n|/(\pi k)}
\label{12}
\end{equation}
where 
\begin{eqnarray}
D_{b} (k) =  \frac{v^2 k^2}{8 \pi} \int_0^1 dx \frac{1}{\sqrt{ m^2 + v^2 k^2 x (1-x)}} \nonumber \\
\times \coth \left( \frac{\sqrt{ m^2 + v^2 k^2 x (1-x)}}{2 T} \right).
\label{14}
\end{eqnarray}

In Fig. \ref{Db} we plot $D_b (k)$ at $T=0$ near and far away from a critical point.
 At the smallest $k$, $D_b (k)$ scales as $k^2$ for any $m >0$. For large $m$, this behavior holds up to $k \sim k_F$, but at small $m$, the $k^2$ dependence 
 only holds for the smallest $k$ crosses over to  $D_b (k) \propto k$ at $v k \geq m$. 

The frequency-dependent pairing interaction $\lambda (\Omega_n)$ is obtained in  the usual way, by interating the gauge field propagator  $\chi_{tr} (k, \Omega_n)$ in the direction along the FS:
\begin{eqnarray}
\lambda (\Omega_n) &=& \frac{\alpha_1 v}{2 \pi^2 } \int_0^{2 k_F} dk \frac{\sqrt{1-(k/2k_F)^2}}{
2 D_{b} (k) + k_F |\Omega_n|/(\pi k)} 
\label{16}
\end{eqnarray}
where, following MS, we introduced another dimensionless parameter
\begin{eqnarray}
\alpha_1 &=& \frac{k_F}{m^{*} v}
\label{15}
\end{eqnarray}
We will see that it plays the role of 
 the dimensionless coupling.
 
Because $D_b (k)$ interpolates between $k$ and $k^2$ dependences, the momentum integral in (\ref{15}) diverges at $\Omega_n =0$ no matter whether the system is at the QCP or away from the QCP (i.e., whether $m=0$ or $m >0$).  This implies that the pairing in the MS model remains quantum-critical even when the system moves away from the actual critical point where SDW order emerges.   
Alternatively speaking, away from the QCP,   magnetic excitations acquire a mass but the pairing gauge boson remains massless.

\section{the MS model as the $\gamma$ model with varying exponent $\gamma$}
\label{sec:3}

As the pairing is generally a low-temperature/low-frequency 
 phenomenon, it is instructive to analyze first 
the form of $\lambda (\Omega_n$ in the MS model, Eq. (\ref{16}), at $T=0$ and in the limit of small $\Omega_n$.  Suppose that the system is in the paramagnetic phase, where $m$ tends to a 
 constant at $T =0$. Using (\ref{18_a}), we obtain at $T=0$, $m = 4\pi m^* v^2 \alpha_2 = 4\pi k_F v \alpha_2/\alpha_1$. 
  For small $\Omega_n$ the momentum integral in (\ref{16}) is determined by 
 $k$ for which $ D_b (k) k/k_F \sim |\Omega_n|$.  These typical $k$ are small when $\Omega_n$ are small.  Substituting the small $k$ expansion of 
 $D_b (k) = v^2 k^2/(8\pi m) $ (see Eq. (\ref{14})) into (\ref{16}) we obtain
\be   
\lambda (\Omega_n) = \left(\frac{\Omega^{\gamma =1/3}_0}{|\Omega|}\right)^{1/3}
\label{17}
\ee
where
\be
\Omega^{\gamma=1/3}_0 = \frac{512\pi^2}{81 \sqrt{3}} E_F \alpha^2_2
\label{18}
\ee
and $E_F = k^2_F/(2 m^*)$. 
The  corresponding self-energy is 
\be
\Sigma (\omega_m) = \frac{3}{2}~\omega_m^{2/3}~ (\Omega^{\gamma=1/3}_0)^{1/3}.
\label{18a}
\ee

This $\lambda (\Omega_m)$ is the same as in the 
$\gamma$-model with $\gamma =1/3$.  
 The onset temperature for the pairing in the ``$1/3$'' model has been calculated in Refs.\cite{nick,emil} and is $T_{p} \approx 3.31 \Omega^{\gamma=1/3}_0$. 
Combing this result with Eq. (\ref{18}), we obtain
 \be
T_p =T^{\gamma=1/3}_p \approx 119~ E_F~ \alpha^2_2 
\label{19}
\ee  
Observe that $T^{\gamma=1/3}_p$ depends on $\alpha_1$ only via $E_F$. 
 
\begin{figure}
\includegraphics[width=3.0 in]{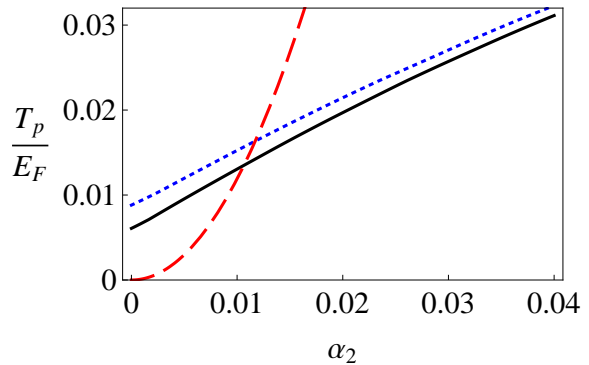}
\caption{
 The actual critical temperature $T_p$  in the MS model with $\alpha_1 = 0.65$
 (the black line) and $\alpha_1 \sim 0.75$ (the dotted blue line)
 (from Ref. \protect\cite{moon})  and 
 $T^{\gamma =1/3}_p$ from  Eq. (\protect\ref{19}) (the dashed red line) 
Although the trend is the same -- both the actual $T_p$ and  $T^{\gamma =1/3}_p$ increase with increasing $\alpha_2$, the functional forms are quite different. The two functions just cross at $\alpha_2 \sim 0.01$.
 Besides, the actual $T_p$ does depend on $\alpha_1$, particularly for small $\alpha_2$, while  $T^{\gamma =1/3}_p$ is independent on $\alpha_1$ (see the text).}
\label{comparison}
\end{figure}

Moon and Sachdev obtained $T_p$ in the MS model numerically, using the 
 full forms of $D_b (k)$ and $m(T)$. In Fig.~\ref{comparison}
 we compare their result with Eq. (\ref{19}).
 We see that while the trend is the same -- $T_p$ icreases with increasing 
$\alpha_2$, the functional forms of the actual $T_p$ and $T^{\gamma=1/3}_p$ 
 are very different, and  Eq. (\ref{19}) agrees with 
 the numerical result for $T_p$ only in a tiny range of $\alpha_2$ near $\alpha_2 \sim 0.01$. Besides,  $T_p/E_F$ obrained numerically does depend on $\alpha_1$, in distinction to Eq. (\ref{19}).  

We looked into this issue more carefully and found that the reason for the 
 discrepancy is that Eq. (\ref{19}) is valid only in a finite range of $\alpha_2$, restricted on both sides, and  the width of this range turns out to be
 vanishingly small for $\alpha_1 \sim 0.5-0.8$ used by MS. 

Specifically, Eq. (19) is obtained under three assumptions: 
\begin{itemize}
\item
 that the $T$-dependence of the bosonic mass can be neglected and
 the bosonic propagator can be evaluated at $T=0$ -- this is justifed if 
$m \approx 4\pi m^* v^2 \alpha_2 >> T^{\gamma=1/3}_p$;
\item
that typical $vk$ are  small compared to $m$, i..e., $D_b(k)$ can be approximated by its small $k$ form
$D_b = v^2 k^2/(8\pi m)$;
\item 
that typical $k$ are small compared to $k_F$.
\end{itemize}
We verified that the first assumption is well satisfied 
 for all $\alpha_1 < 1$ and arbitrary $\alpha_2$ 
 and is not the subject of concern. 

 To verify the second assumption we note that typical $k$ leading to (\ref{19}) are of order $(m k_F \Omega_n/v^2)^{1/3}$, and typical 
$\Omega_m \sim T^{\gamma=1/3}_p$. Substituting the numbers, we find that
 for typical $k$ 
\be
\frac{vk}{m} \sim 2 \alpha_1 n^{1/3}
\label{21}
\ee
 where $n \geq 1$ is a number of a Matsubata frequency (typical $n = O(1)$).
The condition $vk <<m$ is then satisfied when $\alpha_1$ is small enough.  
 There is no dependence on $\alpha_2$ in this equation, hence for
 sufficiently small $\alpha_1$ the condition 
$v k < m$ is satisfied for all $\alpha_2$. 

 The third condition is satisfied at small $\alpha_2$ but breaks out at larger $\alpha_2$. Indeed, for typical  $k$,
\be
\frac{k}{k_F} \sim 27 \alpha_2 n^{1/3}
\label{20}
\ee
 The condition $k/k_F <<1$ is  satisfied when the r.h.s. of (\ref{20}) is small.

\begin{figure}
\includegraphics[width=3.0 in]{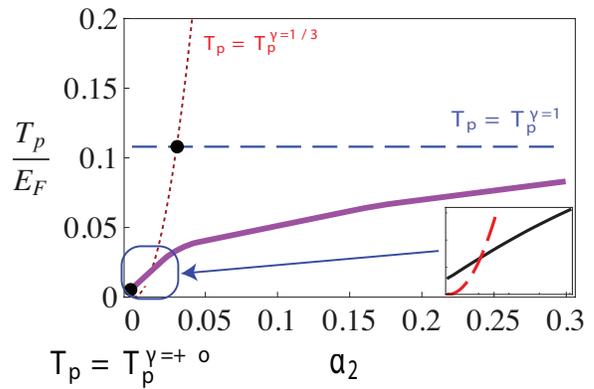}
\caption{
The interplay between $T_p$ in the MS model 
 and in the  $\gamma$-model with the pairing kernel $\lambda (\Omega_n) \propto 1/\Omega^{\gamma}_n$.   
There is a very tiny range near $\alpha_2 \sim 0.01$ where 
$T_p$ in the MS model is the same as in $\gamma =1/3$ model: $T_p = T^{\gamma=1}_p = 119 E_F \alpha^2_2$.  At larger $\alpha_2$, the MS model becomes equivalent to the $\gamma$ model with varying 
$\gamma >1/3$. The  value of $\gamma$ increases with $\alpha_2$ and approaches
$\gamma =1$ at large $\alpha_2$. $T_p$ in this regime deviates from $\alpha^2_2$ dependence of $T^{\gamma =1/3}_p$  and eventually saturates at $T_p = T^{\gamma=1}_p =0.108 E_F$.  At smaller $\alpha_2$,  the MS model becomes equivalent to the $\gamma$ model with varying $\gamma <1/3$.  $T_p$ in this regime again 
deviates from $\alpha^2_2$ dependence of $T^{\gamma =1/3}_p$  and tends to a non-zero value at a magnetic QCP, where $\alpha_2 =0$. 
In this limit, the MS model becomes equivalent to the $\gamma$-model with $\gamma = 0+$ ($\lambda (\Omega_n) \propto \log \Omega_n$), and  $T_p = T^{\gamma=0+}_p \approx 1.8 \frac{E_F}{\alpha_1} e^{-\frac{\pi^2}{2\sqrt{\alpha_1}}}$. }   
\label{schematic}
\end{figure}

\begin{figure}
\includegraphics[width=3.0 in]{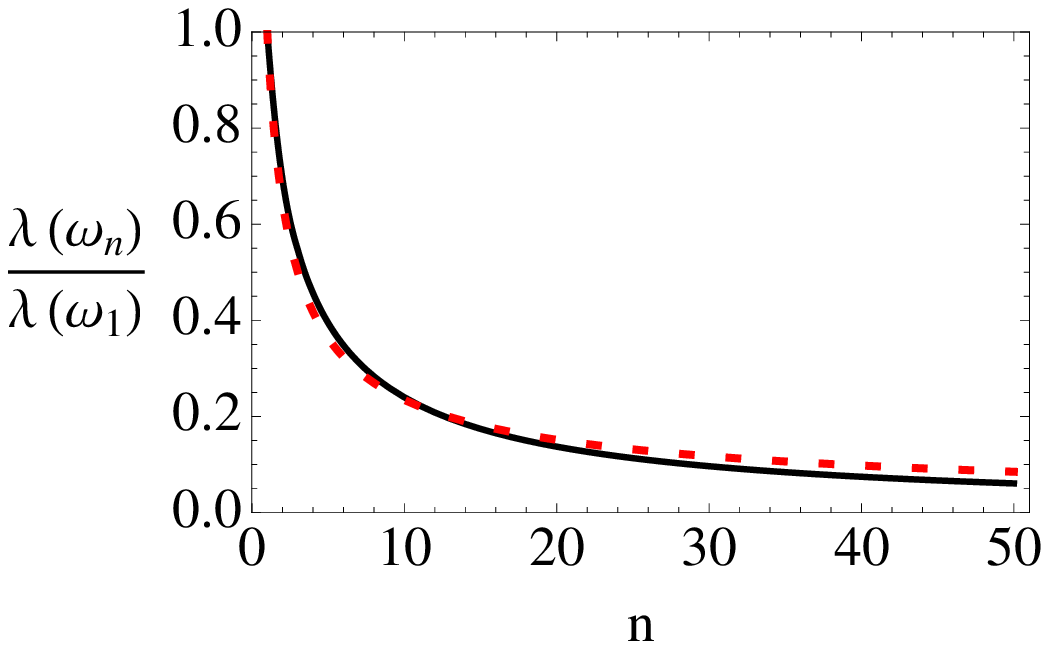}
\includegraphics[width=3.0 in]{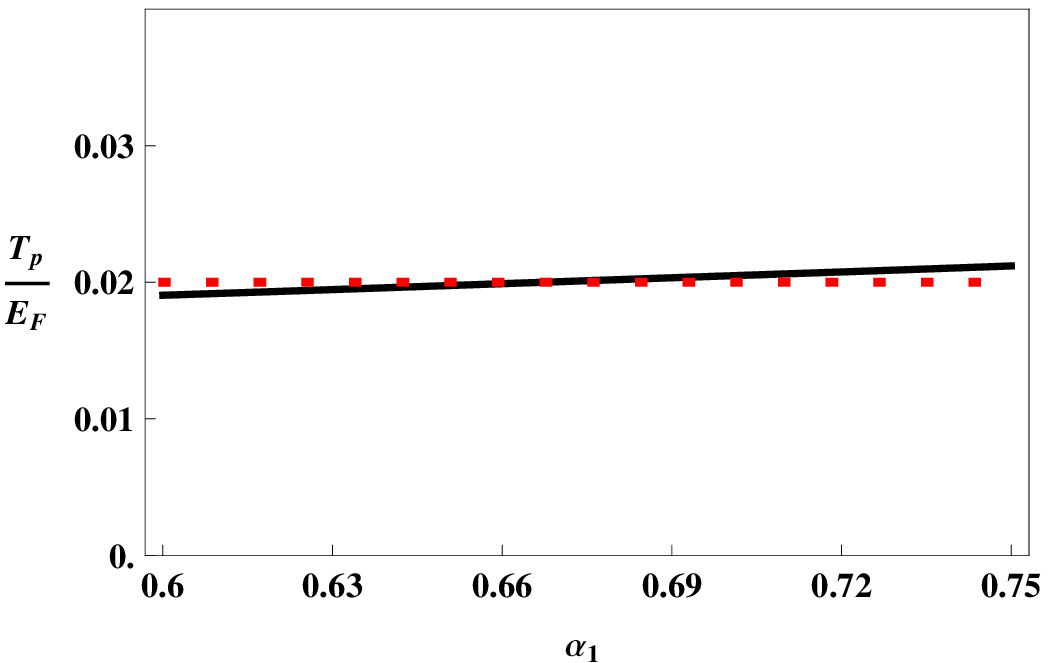}
\caption{ Solid lines: the dimensionless pairing kernel $\lambda (\Omega_n) $ (panel (a)) for
$\alpha^2_1 = 0.32$,
 and the critical temperature $T_p$ vs the  coupling $\alpha_1$ (panel (b))
  at $\alpha_2 = 0.02$ ($\alpha_2$ measures the distance to an 
 antiferromagnetic QCP).
  The dashed (red) line in panel (a) is $\lambda (\Omega_m) = 
(\Omega^{\gamma}_0/\Omega_n)^\gamma$. The best fit corresponds to $\gamma =0.6$ and $\Omega^{\gamma=0.6}_0 \approx 0.045 E_F$.  The dotted (red) 
line in panel (b) is $T_p$ in the $\gamma$ model for $\gamma =0.6$: $T^{\gamma =0.6}_p \approx 0.5 \Omega^{\gamma=0.6}_0 = 0.02 E_F$. The agreement is quite good, but there is some
 variation of the actual $T_p/E_F$ with $\alpha_1$ implying that the value of $\gamma$ in the fit of $\lambda (\Omega_n)$ by $1/\Omega^{\gamma}$ slightly varies with $\alpha_1$.}
\label{coupling002}
\end{figure}

\begin{figure}
\includegraphics[width=3.0 in]{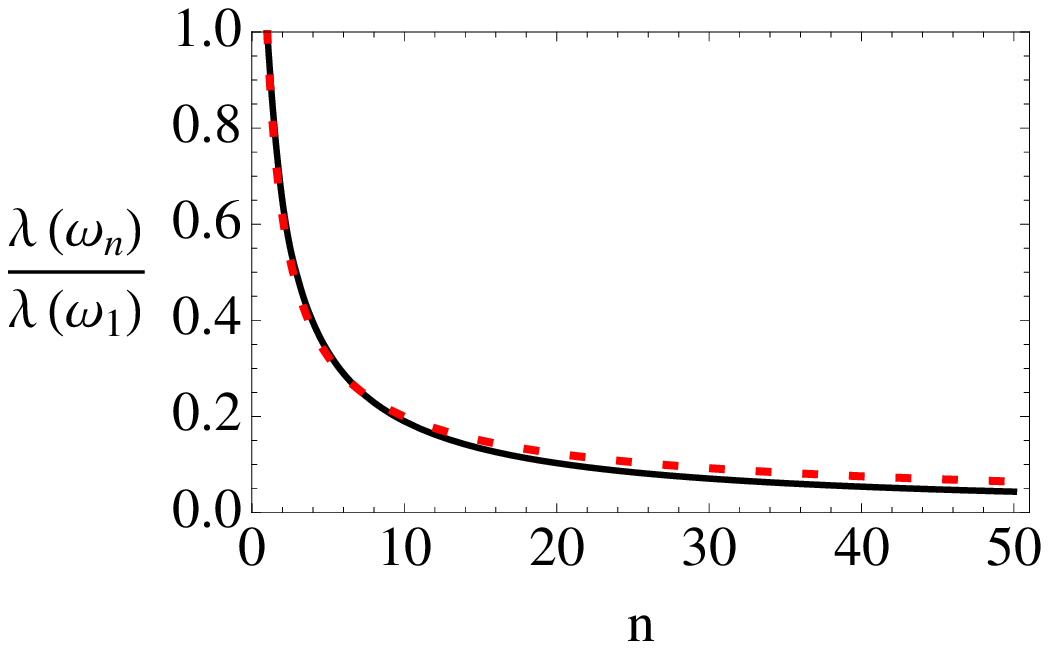}
\includegraphics[width=3.0 in]{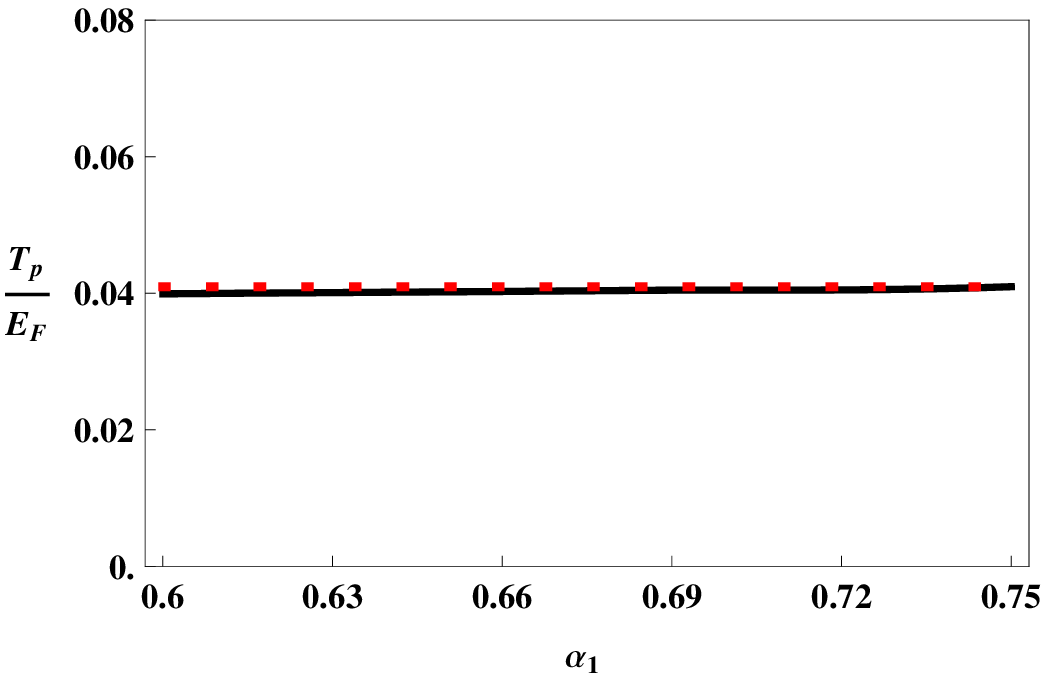}
\caption{Same as in Fig.  \ref{coupling002}  but for a larger $\alpha_2 =0.06$.
 The best fit by a power-law (dashed (red) line in panel (a)) 
 now corresponds to   $\gamma =0.7$, and $\Omega^{\gamma=0.7}_0 \approx 1.11 E_F$. The dotted (red) line is  $T^{\gamma =0.7}_p \approx 0.38 \Omega^{\gamma=0.7}_0 \approx 0.042 E_F$. The areement is quite good, and there is less variation of $t_P/E_F$ than in Fig.  \ref{coupling002}.}
\label{coupling006}
\end{figure}

\begin{figure}
\includegraphics[width=3.0 in]{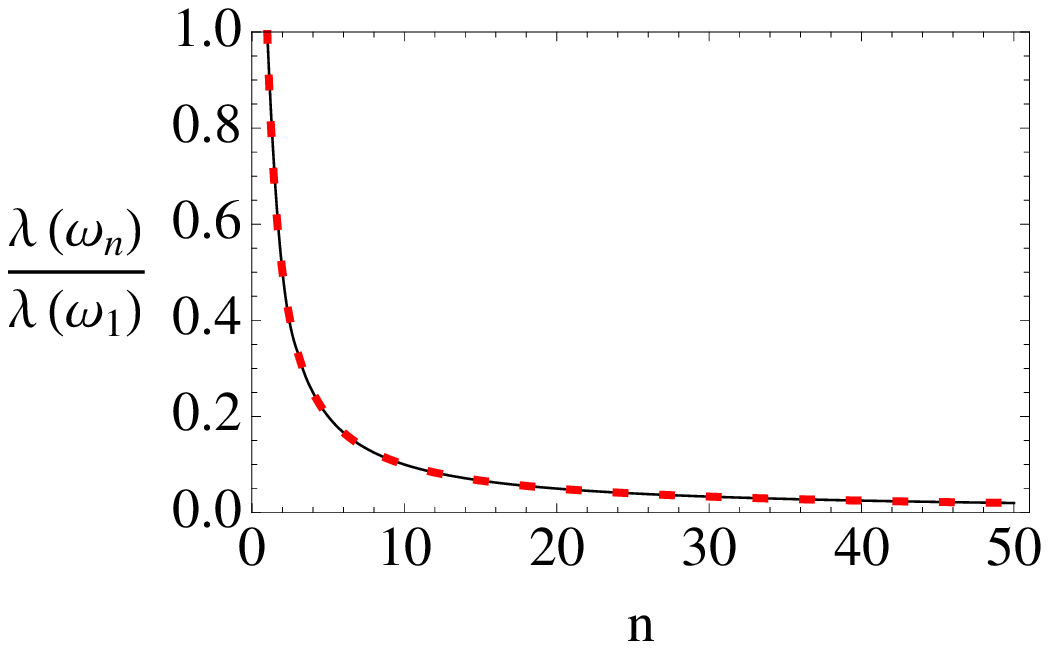}
\includegraphics[width=3.0 in]{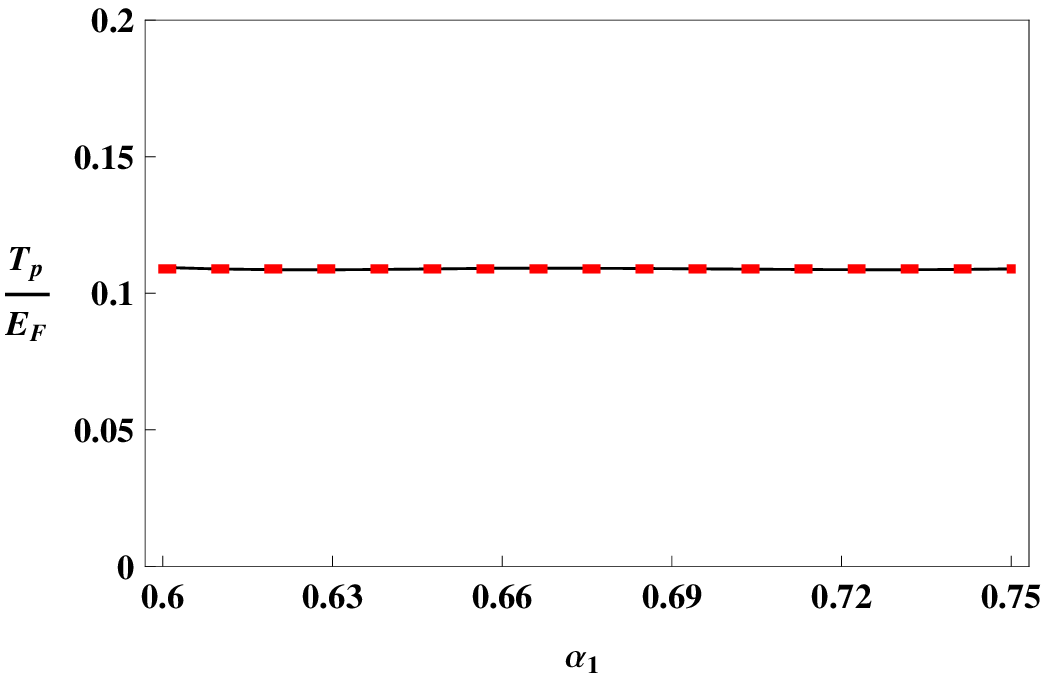}
\caption{Dimensionless coupling constant $\lambda_T (\omega_n) $ and critical temperature in the extreme limit of very large $\alpha_2 = 600$. 
Details are the same as the Fig. \ref{coupling002}. 
The power-law exponent in the fit $\gamma =1$, $\Omega^{\gamma=1}_0 = 4E_F/(3\pi) =0.424 E_F$, and $T^{\gamma =1}_p \approx 0.254 \Omega^{\gamma=1}_0 \approx 0.108 E_F$.This $T^{\gamma =1}_p$ perfectly matches $T_p$ for all $\alpha_1$.}
\label{coupling600}
\end{figure}

\subsection{crossover at ``large'' $\alpha_2$}

 The large numerical prefactor in (\ref{20}) implies that Eq. (\ref{19}) becomes invalid  beginning from quite small $\alpha_2 \sim 10^{-2}$. Once typical $k$ become larger than $k_F$, $\lambda (\Omega_n)$ has to be re-evaluated because 
 the momentum integral in (\ref{16}) only extends up to $k_F$. We assume and then verify that in this situation the Landau damping term $\Omega_n/k$ is larger than $D_b (k)$ for 
  all $k < k_F$ and for $\Omega_m$ comparable to the actual $T_p$, which differs from  $T^{\gamma=1/3}_p$ in Eq. (\ref{19}). Neglecting $D_b(k)$ in (\ref{16}) we obtain in this situation
 \be
\lambda (\Omega_n) = \frac{\Omega^{\gamma=1}_0}{ |\Omega_n|}
\label{25}    
\ee
where $\Omega^{\gamma=1}_0 = 4E_F/(3\pi)$.  Observe that $\Omega^{\gamma=1}_0$ does not depend on $\alpha_2$, in distinction to $\Omega^{\gamma =1/3}_0$ in Eq. (\ref{18}). 

This form of $\lambda (\Omega_n)$ corresponds to the 
 $\gamma-$ model with $\gamma=1$. $T_p$ in $\gamma=1$ model has been obtained in \cite{ch_sch,emil}: 
\be
T^{\gamma=1}_p = 0.254 \Omega^{\gamma=1}_0 = 0.108 E_F.
\label{26}
\ee
 This should be the actual $T_p$ in the MS model at large $\alpha_2$. 
 Comparing this $T^{\gamma=1}_p$ with $T^{\gamma=1/3}_p$ from Eq. (\ref{19}),
 we clearly see the difference: while $T^{\gamma=1/3}_p$ scales as $\alpha^2_2$, the actual $T_p$ saturates at large $\alpha_2$.   

 In the crossover regime,
  the coupling $\lambda (\Omega_n)$ and the pairing instability temperature $T_p$ should interplolate between $\gamma=1/3$ and $\gamma=1$ behavior. 
 The two temperatures $T^{\gamma=1/3}_p$ and $T^{\gamma=1}_p$ cross at 
$\alpha_2 \sim 0.03$, and the crossover should be around these $\alpha_2$.
  We show the crossover in Fig.\ref{schematic}.
 Note that $\alpha_2 \sim 0.03$ is consistent with  our earlier easimate of 
$\alpha_2 \sim 10^{-2}$ at which  typical $k$ becomes comparable to $k_F$.

To understand the system behavior in the crossover regime,
  we computed $\lambda (\Omega_n)$ 
 for several $\alpha_2$  and found that 
  for each given $\alpha_2$,  $\lambda (\Omega_n)$ can be well fitted over a wide frequency range  by $\Omega^{-\gamma}_n$ form, where $\gamma$ changes 
between $1/3$ and $1$ when $\alpha_2$ increases. In other words, for a given $\alpha_2$, the MS model is equivalent to a $\gamma-$model with a particular exponent $\gamma$.  
In Figs . \ref{coupling002} and \ref{coupling006}, we plot $\lambda (\Omega_n)$ vs the number of the Matsubara frequency, and the actual $T_p$ as a function of $\alpha_1$ for $\alpha_2 =0.02$ and $\alpha_2 =0.06$. The best fits to $\gamma$-model correspons to $\gamma=0.6$, $\Omega^{\gamma=0.6}_0 = 0.04$,
 and $\gamma=0.7$, $\Omega^{\gamma=0.7}_0 = 0.11$, respectively.
 The dashed lines in the figures for $T_p$ are the values of the pairing instability temperature in the $\gamma-$model (Ref.\cite{emil}): $T^{\gamma=0.6}_p \approx 0.5 \Omega^{\gamma=0.6}_0 = 0.02 E_F$ and $T^{\gamma=0.7}_p \approx 0.38  \Omega^{\gamma=0.7}_0 = 0.042 E_F$. We see that the agreement is nearly perfect, although for $\alpha_2 =0.02$, there is some residual dependence of $T_p$ on $\alpha_1$ which is not present in the $\gamma-$model.  In Fig.  \ref{coupling600} we show the fit to $1/\Omega_n$ from of $\lambda_n$ and $T_p$ for the limiting case of large $\alpha_2$. We see that the actual $T_p$ perfectly 
matches $T^{\gamma=1}_p = 0.108 E_F$, and the ratio $T_p/E_F$ is totally independent on $\alpha_1$.

\subsection{crossover at small $\alpha_2$}

We next consider what happens at small $\alpha_2$, when typical $k$ are much smaller than $k_F$. At the first glance, the analogy with $\gamma =1/3$ model should work in the small $\alpha_2$ range because all three conditions used in the derivation of Eqs. (\ref{17})-(\ref{19}) are satisfied.  If so, $T_p$ should scale as $\alpha^2_2$ and vanish at $\alpha_2 =0$, see Eq. (\ref{19}).
 This, however, is not the case as is clearly seen from Fig.\ref{comparison} -- the actual $T_p$ remains nonzero at $\alpha_2 =0$.  

The reason for this discrepancy is that the validity 
 of the three conditions $m >>T, vk << m$, and $k << k_F$ 
 in fact only implies that Eq. (\ref{19}) is self-consistent 
at small $\alpha_2$. However, there may be another contribution to $T_p$ from 
 the range $vk >m$, and if such additional contribution does exist, it may 
oversahadow the contribution from 
$v k <m$. A good example of such behavior is the case of phonon superconductors at strong coupling~\cite{phon_rev,ad,combescot}: McMillan $T^{MM}_p \sim \omega_D e ^{-(1+\lambda)/\lambda}$ is self-consistently obtained as a contribution from fermions in a Fermi-liquid region
 of $\omega < \omega_D$. Yet, there is another contribution to $T_p$ from fermions outside of the Fermi-liquid regime, and at strong coupling ($\lambda >>1$) this second contribution overshadows the contribution from the Fermi liquid range and yields Allen-Dynes expression~\cite{ad} 
$T^{AD}_p \sim \omega_D \sqrt{\lambda} \gg T^{MM}_p$.  

 To verify whether the same happens in our case, consider the limit 
 $\alpha_2=0$, when $T_p$ given by (\ref{19}) vanishes.  If the actual 
 $T_p$ remains finite at this point, it necessary comes 
 from $v k >m$ simply because $m(T=0) =0$. Re-evaluating  $\alpha (\Omega_n)$ at $\alpha_2 =0$ and $T=0$ we obtain $D_b (k) = v k/8$ and  
\be
\lambda (\Omega_n) = \frac{2 \alpha_1}{\pi^2} \left[\sqrt{1 + \frac{|\Omega_n|}{\pi v k_F}} \tanh^{-1}\left(\frac{1}{\sqrt{1 + \frac{|\Omega_n|}{\pi v k_F}}}\right) -1 \right]\label{22}
\ee
At small $\Omega_n <<  v k_F$ this reduces, with logarithmical accuracy,  to 
\bea
\lambda (\Omega_n) &=& \frac{\alpha_1}{\pi^2} \log{\frac{4\pi k_F v}{e^2 |\Omega_n|}} = \frac{\alpha_1}{\pi^2}  \log{\left(\frac{\Omega^{\gamma=0+}_0}{\Omega_n}\right)}, \nonumber \\
&&~~~{\text where} ~~~\Omega^{\gamma=0+}_0 = \frac{8 \pi E_F}{e^2 \alpha_1}
\label{23}
\eea
The model with such $\lambda (\Omega_n)$ is again equivalent to the $\gamma$-model, but with $\gamma = 0+$. The $\gamma = 0+$ model with logarithmical kernel 
has been analysed by Son~\cite{son} and others.~\cite{bedell,ch_sch,pre_exp}
 It {\it does have} a pairing instability at a non-zero temperature, i.e., the actual $T_p$ does not vanish 
at $\alpha_2=0$, in distinction to Eq. (\ref{19}). 
 For $\alpha_1 = 0.6$, we found $T^{\gamma =0+}_p \sim 0.005 E_F$.  
 This obviously implies that Eq. (\ref{19}) is valid only in limited range of $\alpha_2$, bounded from both ends. Furthermore, compairing 
 this $T^{\gamma=0+}_p \sim 0.005 E_F$ with $T^{\gamma=1/3}_p$ from Eq. (\ref{19}), we find that they become comparable at the same $\alpha_2 \sim 10^{-2}$, where  $T^{\gamma=1/3}_p$ becomes comparable to  $T^{\gamma=1}_p$. As the consequence, the width of the range where Eq. (\ref{19}) is valid turns out to be
 vanishingly small number-wise, although  paramer-wise it is  $O(1)$.
This explains why the actual $T_p$ is so different from  $T^{\gamma=1/3}_p$
 (see Fig. \ref{comparison}). 
 
 We will discuss the form of $T^{\gamma=0+}_p$  in more detail in the next section and here only note that the total $T_p$ in the MS model is the sum of 
 contributions from $v k >m$ and $v k <m$, i.e., $T_p \sim  T^{\gamma = 0+}_p + T^{\gamma =1/3}_p$.  In the crossover range where $T^{\gamma = 0+}_p$ and $T^{\gamma = 1/3}_p$ are comparable, the MS model is  most likely again fitted 
 by the $\gamma$-model with varying $\gamma$  between $0$ and $1/3$ (i.e., by 
different $\gamma$ for different $\alpha_2$).
  However, the width of the crossover range at small $\alpha_2$ is narrow,
 and we didn't attempt to fit the data at the smallest $\alpha_2$ by the $\gamma-$model with $0<\gamma <1/3$. 

\section{the MS model at the SDW QCP -- the equivalence with color superconductivity}
\label{sec:4}

The model with the logarithmical pairing kernel, as in Eq. (\ref{23}),
 is most known as a model for color superconductivity of quarks due to the exchange by gluons. 
For small $\omega_n$, the fermionic self-energy for logarithmical $\lambda (\Omega_n)$ given by (\ref{23}) is
\be
\Sigma (\omega_n) =\int_0^\omega d \omega' \lambda (\omega') = g \omega \log{\left( \frac{\Omega^{\gamma =0+}_0}{\omega_n} \right)} .
\label{23_a}
\ee
To shorten notations, we define $g = \alpha_1/\pi^2$. The pairing kernel at small frequencies is $\lambda (\Omega_m) = g \log{\left( \frac{\Omega^{\gamma =0+}_0}{\omega_n} \right)}$. 

The equation for the pairing vertex is, to the logarithmical accuracy, 
\begin{eqnarray}
\Phi(\omega) &=& \frac{g}{2} \int_{T^{\gamma = 0+}_p}^{\Omega^{\gamma =0+}_0} d \omega' \frac{\Phi(\omega')}{\omega'} \frac{{\rm log}(\frac{\Omega^{\gamma =0+}_0}{| \omega'-\omega |}) + {\rm log}(\frac{\Omega^{\gamma =0+}_0}{| \omega'+\omega |}) }{1+ g \, {\rm log}\left( \frac{\Omega^{\gamma =0+}_0}{\omega'} \right)} \nonumber \\
& = & g \int_{T_c}^{\Omega^{\gamma =0+}_0} d \omega' \frac{\Phi(\omega')}{\omega'} \frac{{\rm log}(\frac{\Omega^{\gamma =0+}_0}{\sqrt{| \omega'^2-\omega^2 |}}) }{1+g \, {\rm log}\left( \frac{\Omega^{\gamma =0+}_0}{\omega'} \right)} .
\label{pairing}
\end{eqnarray}
The logarithm in the denominator is the contribution from the self-energy, Eq. (\ref{23_a}).
   At small coupling, $g <<1$, which we only consider, the self-energy can be dropped, and the condition that the solution of (\ref{pairing}) exists reduces to $g \log^2 (\Omega^{\gamma =0+}_0/T^{\gamma = 0+}_p) = O(1)$, i.e., $\log T_c \propto 1/\sqrt{g}$.
  A more accurate solution was obtained by Son~\cite{son}:
\be
T^{\gamma=0+}_p = c~ \Omega^{\gamma =0+}_0 ~e^{-\frac{\pi}{2\sqrt{g}}} = 
 {\bar c}~ \frac{E_F}{\alpha_1}~ e^{-\frac{\pi^2}{2\sqrt{\alpha_1}}} 
\label{24}
\ee     
where $c$ and ${\bar c} = 8 \pi c/e^2$ are  numbers $O(1)$. 
 The smaller is $\alpha_1$, the smaller is $T^{\gamma = 0+}_p$ and the smaller is the crossover scale below which the actual $T_p$ deviates from $T^{\gamma=1/3}_p$.

\begin{figure}
\includegraphics[width=3.0 in]{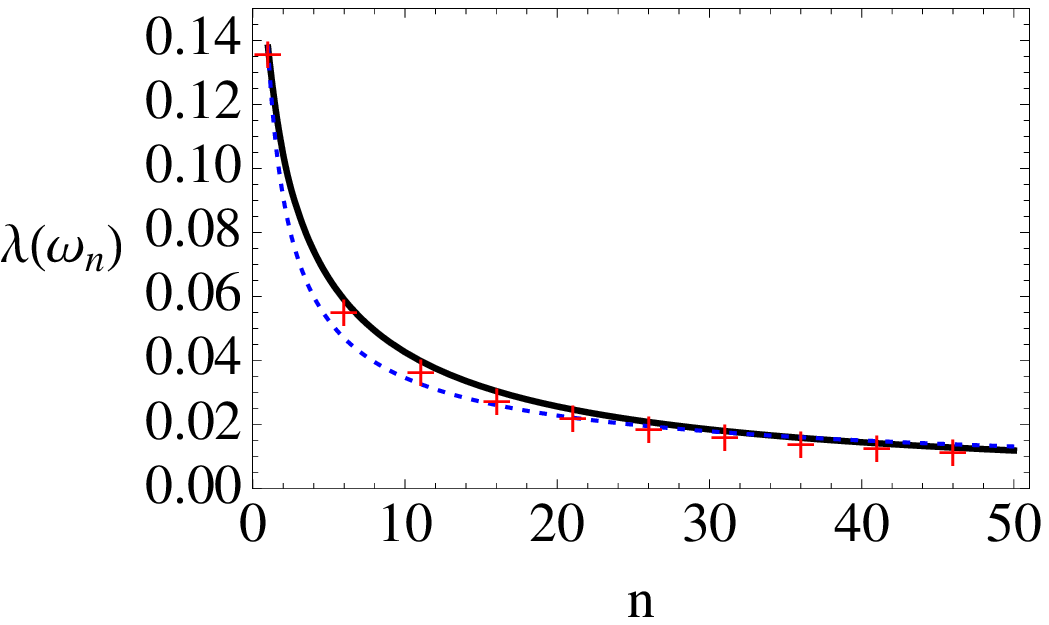}
\includegraphics[width=3.0 in]{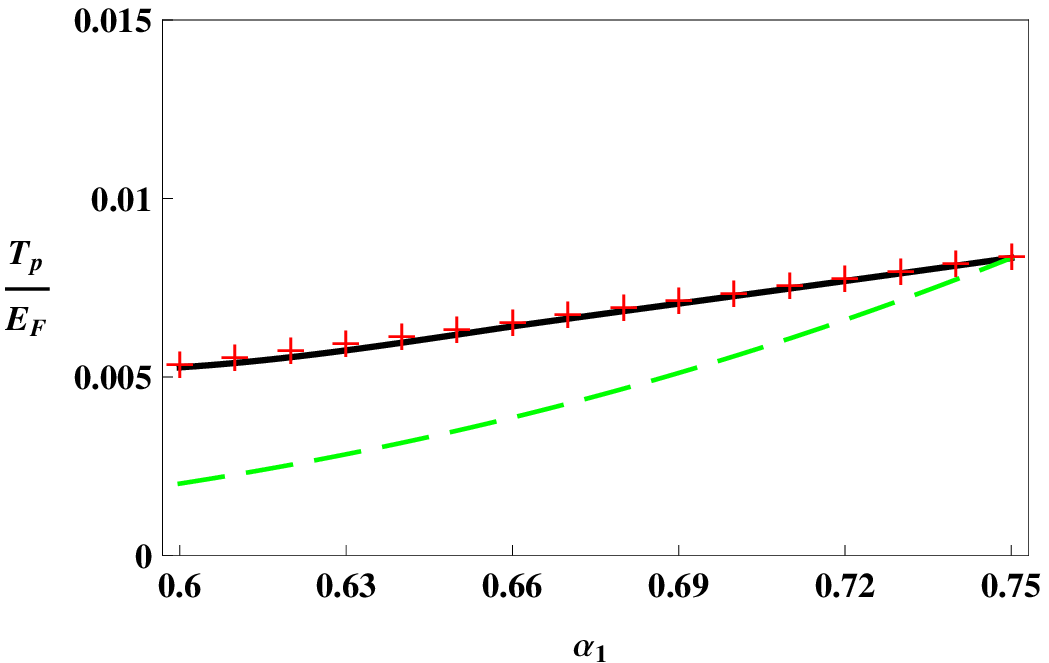}
\caption{Solid lines: 
 the pairing kernel $\lambda (\Omega_n)$ (panel (a)) and the 
critical temperature $T_p$ in the MS model at the critical point, $\alpha_2 = 0$ (panel (b)). 
 In panel (a) we set $\alpha^2_1 =0.32$. The crossed (red)
 line in panel (a) is the fit to 
$\lambda (\Omega_n) = (\alpha_1/\pi^2) \log (\Omega^{\gamma=0+}_0/\Omega_n)$ with
$\Omega^{\gamma=0+}_0 = \frac{8 \pi E_F}{e^2 \alpha_1}$ 
(see Eq. (\protect\ref{23})). For comparison, the dotted (blue) line is the fit by a power-law with 
 $\gamma =0.6$. Obviously, the fit by $\gamma = 0+$ form is much better.
The crossed (red) line in panel (b) corresponds to $T^{\gamma= 0+}_p \approx 1.8 
\frac{E_F}{\alpha_1} e^{-\frac{\pi^2}{2\sqrt{\alpha_1}}}$. The fit is almost perfect (solid and dashed lines are virtually undistinguishable). For comparison, we also show the fit to the BCS behavior $T^{BCS}_p \propto 1/\alpha_1 e^{-\pi/2 \alpha_1}$ (the dashed (green) line), adjusting the prefactor to match $T_p$
 at $\alpha_1 =0.75$. We clearly see that $T_p$ in the MS model follows
  $T^{\gamma= 0+}_p$ rather than the BCS form.}
\label{coupling000}
\end{figure}
 
To obtain the prefactor $c$ in (\ref{24}), one has to go beyond the 
 leading logarithmical accuracy and  include (i) 
$\sqrt{g}$ corrections to the argument of the exponent and (ii) the 
 actual soft high-energy cutoff in the full expression for $\lambda (\Omega_n)$, see Eq. (\ref{22}). 
The $\sqrt{g}$ corrections to the argument of the exponent come from 
fermionic self-energy.  These corrections have been computed analytically in 
 Ref.~\cite{pre_exp}, and the result is that the fermionic self-energy contributes the universal factor $e^{(\pi^2 +4)/8} = 0.177$ to $c$.
The effect of a soft high-energy cutoff depends on how the logarithmical form of $\lambda (\Omega_n)$ is cut and is model-dependent.

To the best of our knowledge, the non-trivial dependence of $T^{\gamma=0+}_p$ 
on the coupling $g =\alpha_1/\pi^2$  has not been verified numerically. 
 The equivalence between the MS model at $\alpha_2=0$ and the $\gamma$ model with $\gamma = 0+$ allows us to do this and also to determine the value of the prefactor in our case.  In Fig. (\ref{coupling000}) we show the fit of $\lambda (\Omega_n)$ for $\alpha_2=0$ to $1/\Omega_n$ dependence (which is almost perfect) and the numerical results for $T_p/E_F$ vs $\alpha_1$. We fitted the data by  Eq. (\ref{24}), i.e. by $(1/\alpha_1)~e^{-\pi^2/(2\sqrt{\alpha_1})}$ dependence on $\alpha_1$, and, for comparison, by BCS dependence.  
We see that the fit by Eq. (\ref{24}) is nearly perfect. We emphasize that $1/\sqrt{\alpha_1}$ dependence in the exponent and the presence of $1/\alpha_1$ in the overall factor are both relevant to the success of the fit.  
 We view this agreement as the solid verification of Son's formula for color superconductivity.  From the fit we also determined $c \approx 0.037$, and ${\bar c} \approx 1.8$. 
  
\section{Conclusions} 
\label{sec:5}

To conclude, in this paper we analysed the onset temperature $T_p$ 
for the pairing in cuprate superconductors  at small doping, when tendency towards antiferromagnetism is strong.
 We considered the model, introduced by Moon and Sachdev, which assumes that the Fermi surface retains pocket-like form, same as in an antiferromagnetically ordered state,  even when long-range antiferromagnetic order disappers.
 Within this model, the pairing between fermions is mediated by a gauge boson,
 whose propagator remains massless despite that magnon dispersion
 acquires a finite gap in the magnetically disordered  phase. 
From the point of view of quantum-critical phenomenon, the pairing problem then remains quantum critical in the sense that that the pairing is mediated by a gapless boson, and the same gapless boson that mediates pairing also accounts for the destruction of a Fermi-liquid behavior 
 down to the smallest frequencies.

We related the MS model to the generic $\gamma-$model of quantum-critical pairing with the pairing kernel  $\lambda (\Omega_n) \propto 1/\Omega^{\gamma}_n$
 and the corresponding normal state self-energy $\Sigma (\omega_n) \propto \omega^{1-\gamma}_n$.  We showed that, over some range of parameters,
 the pairing problem in the MS model in a paramegnetic phase predominantly 
comes from fermions with momenta $k << k_F$,  where $k_F$ is the Fermi momentum,, and $v k << m$, where $v$ is the Fermi velocity, and $m$ is the magnon gap. 
 In this situation, the pairing in the MS model 
is equivalent to that in the $\gamma$-model with $\gamma =1/3$
 We found that $T_p$ scales as $\alpha^2_2$, where $\alpha_2$ is the parameter which measures how far the system moves away from  the SDW phase.

On a more careful look, we found that the range of $\alpha_2$ where the analogy with the $\gamma =1/3$ model works is bounded at both small and 
``large'' $\alpha_2$. At ``large'' $\alpha_2$  (which numerically are still
 quite small), the condition $k << k_F$ breaks down. We demostrated that, in this situation, the MS model becomes equivalent to $\gamma$ model with varying $\gamma$, whose value 
increases with  $\alpha_2$ and approaches $\gamma =1$ at large $\alpha_2$. 
 In the crossover regime, the actual $T_p (\alpha_2)$ 
deviates from the $\alpha^2_2$ dependence and eventually saturates at large $\alpha_2$ at $T_p \sim 0.1 E_F$. 

At the smallest $\alpha_2$, we found that the analogy with $\gamma =1/3$ model again bveaks down, this time because the contribution to $T_p$ from $v k << m$
 gets overshadowed by the contribution 
  from $v k >>m$.  In this situation, the MS model becomes equivalent to the 
$\gamma$-model with $\gamma <1/3$. Finally, at $\alpha_2 =0$, i.e., at the SDW 
 transition point,
 the MS model becomes equivalent to the $\gamma$ model with $\gamma = 0+$, for 
which $\lambda (\Omega_m) \propto \log \Omega_m$.
 
The $\gamma = 0+$ model has been studied in the context of color superconductivity and superconductivity near 3D ferro- and antiferromagnetic QCP. The pairing problem with the logarithmical pairing kernel is similar to the 
BCS theory in the sense that the  fermionic self-energy can be neglected in the leading logarithmic approximation at weak coupling $g$. However, in distinction to the BCS theory, $\log T_c$ scales 
 not as $1/g$ but as $1/\sqrt{g}$.  We used the analogy between the MS and the 
$\gamma= 0+$ model at small frequencies and the fact that $\lambda (\Omega_n)$ in the MS model does converge at high frequencies (i.e., one does not have to introduce a cutoff), and computed $T_p$ numerically. We explicitly verified the $1/\sqrt{g}$ behavior in the exponent and also verified the $1/g$ dependence of the overall factor for $T_p$. 
 To the best of our knowledge, this has not been done before. 

 Overall, our results imply that the MS model is  quite rich. It 
 contains the physics of the quantum-critical pairing in all 
 models with kernels $\lambda (\Omega_n) \propto 1/\Omega^\gamma_n$ and
 $0<\gamma \leq 1$. In addition, the MS model  taken at the point of the
 magnetic instability is equivalent to the 
   model for color superconductivity. 
It is amaizing that one can get useful information about color superconductivity by studying the pairing at the antiferromagnetic instability in the cuprates.    
We thank S. Sachdev, M. Metlitskii and J. Schmalian for 
 the interest to this work and useful comments. This work was supported by 
  NSF DMR-0757145 and the Samsung Scholarship (E.G. Moon) and by 
 NSF-DMR-0906953 (A. V. Ch.)

\end{document}